\documentclass{elsart1p}
\usepackage{ifpdf}
\usepackage{graphicx,amssymb,lineno}
\ifpdf
\usepackage[%
  pdftitle={The SINFONI pipeline},%
  pdfauthor={Andrea Modigliani},%
  pdfsubject={The preprint document class elsart},%
  pdfkeywords={instructions for use, elsart, document class},%
  pdfstartview=FitH,%
  bookmarks=true,%
  bookmarksopen=true,%
  breaklinks=true,%
  colorlinks=true,%
  linkcolor=blue,anchorcolor=blue,%
  citecolor=blue,filecolor=blue,%
  menucolor=blue,pagecolor=blue,%
  urlcolor=blue]{hyperref}
\else
\usepackage[%
  breaklinks=true,%
  colorlinks=true,%
  linkcolor=blue,anchorcolor=blue,%
  citecolor=blue,filecolor=blue,%
  menucolor=blue,pagecolor=blue,%
  urlcolor=blue]{hyperref}
\fi

\makeatletter
\def\elsartstyle{%
    \def\normalsize{\@setfontsize\normalsize\@xiipt{14.5}}
    \def\small{\@setfontsize\small\@xipt{13.6}}
    \let\footnotesize=\small
    \def\large{\@setfontsize\large\@xivpt{18}}
    \def\Large{\@setfontsize\Large\@xviipt{22}}
    \skip\@mpfootins = 18\p@ \@plus 2\p@
    \normalsize
}
\@ifundefined{square}{}{\let\Box\square}
\makeatother

\begin{document}

\begin{frontmatter}
\title{The SINFONI pipeline}

\author[Garching]{Andrea Modigliani\corauthref{cor}},
\corauth[cor]{Corresponding author.}
\ead{Andrea.Modigliani@eso.org}
\author[Garching]{Wolfgang Hummel},
\author[Garching,MPE]{Roberto Abuter},
\author[Paranal]{Paola Amico},
\author[Garching]{Pascal Ballester},
\author[MPE]{Richard Davies},
\author[Paranal]{Christophe Dumas},
\author[NED]{Matthew Horrobin},
\author[Garching]{Mark Neeser},
\author[Garching]{Markus Kissler-Patig},
\author[Garching]{Mich\`ele Peron},
\author[Leiden]{Juha Reunanen},
\author[MPA]{Juergen Schreiber},
\author[Paranal]{Thomas Szeifert}.

\address[Garching]{ESO, Karl-Schwarzschild-Str. 2, D-85748 Garching, Germany} 
\address[MPE]{Max-Planck-Institut  f\"ur extraterrestrische Physik, Giessenbachstra\ss e, D-85748 Garching, Germany}
\address[Paranal]{ESO, Alonso de Cordova 3107, Vitacura, Casilla 19001, Santiago 19, Chile}
\address[NED]{Astronomical Institute \u201cAnton Pannekoek\u201d, Kruislaan 403, NL-1098 SJ Amsterdam, The Netherlands}
\address[Leiden]{Leiden Observatory, P.O. Box 9513, NL-2300RA Leiden, The Netherlands}
\address[MPA]{Max-Planck-Institut fur Astronomie, Koenigstuhl 17, D-69117 Heidelberg, Germany}


\begin{abstract}
The SINFONI data reduction pipeline, as part of the ESO-VLT Data Flow
System, provides recipes for Paranal Science Operations, and for Data Flow 
Operations at Garching headquarters. At Paranal, it is used for the 
quick-look data 
evaluation. For Data Flow Operations, it fulfills several 
functions: 
creating master calibrations; monitoring instrument health and data quality; 
and reducing science data for delivery to service mode users.
The pipeline is available to the science community for
reprocessing data with personalised reduction strategies and 
parameters. The pipeline recipes can be executed either with EsoRex 
at the command line level or through the Gasgano graphical user interface.
The recipes are implemented with the ESO Common Pipeline Library (CPL).

SINFONI is the Spectrograph for INtegral Field Observations in the Near
Infrared (1.1-2.45 $\mu$m) at the ESO-VLT. SINFONI was developed and build by  
ESO and MPE in collaboration with NOVA.
It consists of the
SPIFFI integral field spectrograph and an adaptive optics 
module which allows diffraction limited and seeing limited observations.
The image slicer of SPIFFI chops the SINFONI field of view on the 
sky in 32 slices which are re-arranged to a pseudo slit. The latter is 
dispersed by one of the four possible gratings (J, H, K, H+K). The detector 
thus sees a spatial dimension (along the pseudo-slit) and a spectral dimension.

We describe in this paper the main data reduction procedures of the SINFONI
pipeline, which is based on SPRED - the SPIFFI data reduction software 
developed by MPE,  and the most recent developments after more than a year of 
SINFONI operations.
\end{abstract}

\end{frontmatter}

\section{Introduction}
\label{intro}

SINFONI is the Spectrograph for Integral Field Observations in the Near 
Infra Red (NIR). It is attached to the Cassegrain focus of Yepun, the fourth 
Unit Telescope (UT4) on Cerro Paranal in northern Chile.
It allows diffraction and seeing limited integral field observations in the 
near infrared (1.1-2.45 $\mu$m) at 3 spatial resolutions: 250, 100 and 25 
mas/pixel. 

We will provide a general introduction to instrument pipelines, then present 
the features of the Common Pipeline Library (CPL) and its front end 
applications Gasgano and EsoRex. These will be followed by an overview of the 
instrument and an introduction to the principle of integral field spectroscopy.
Next we will describe the data reduction cascade and the main data reduction 
challenges. 
Finally we will highlight some scientific results based on pipeline reduced 
data products which were obtained during the science verification phase.

\section{Instrument Pipelines Objectives}
\label{overview}

Today twelve instruments are operational at Paranal. They cover the 
observational domain over a wide range of spectral and spatial resolutions.
ESO uses instrument data reduction pipelines to support the complex 
end to end data flow operation model. The main objectives of the pipeline are 
to:


\begin{itemize}
  \item Support the Garching Data Flow Operation (DFO) department in the:
  \begin{itemize}
    \item Processing of raw calibration frames into master calibrations.

    \item Generation of quality control (QC1) parameters to monitor telescope, 
    instrument and detector performance, so that the quality of 
    calibration and science data can be assessed in due time.

    \item Reduction of sets of raw science data files to science data products.
  \end{itemize}
  \item Support the Paranal Science Operations (PSO) with a quick-look 
  tool to assess the instrument health and the quality of the science and 
  calibration data in due time.

  \item Allow users to perform a personalised data reduction with user
  selected reduction strategies and parameter settings.
\end{itemize}

\noindent
The VLT instrument pipeline releases are publicly available at 
\href{http://www.eso.org/pipelines}{\texttt{www.eso.org/pipelines}}.
More details are given in \cite{ballester06} and \cite{silva05}.

\begin{figure*}[ht]
\centering
\includegraphics*[width=13.5truecm]{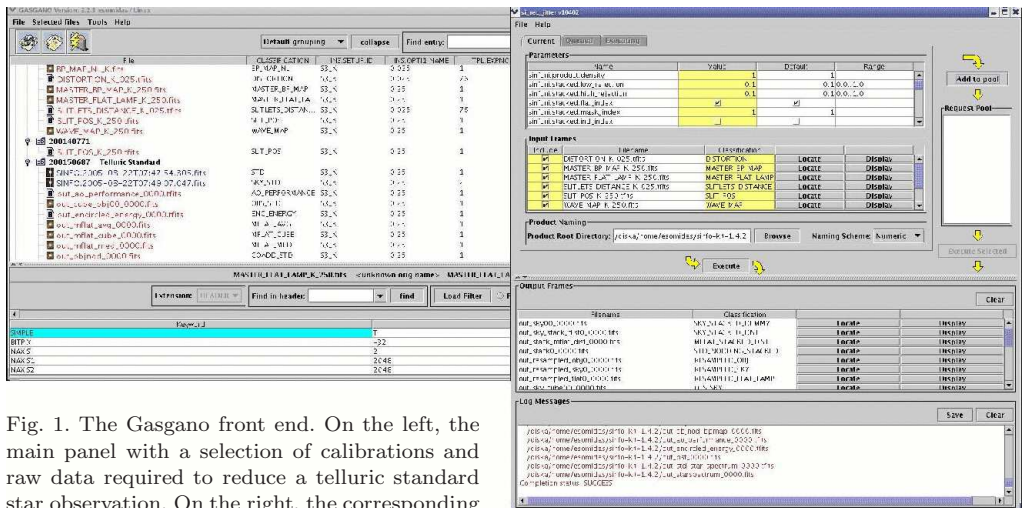}

\vspace{-2truecm}
\begin{minipage}[t]{6.3truecm}
\caption{The Gasgano front end. On the left, the main panel with a 
selection of calibrations and raw data required to reduce a telluric standard 
star observation. On the right, the corresponding recipe execution control 
panel\label{fig:Gasgano}.}
\end{minipage}
\hfill
\begin{minipage}[t]{6.7truecm}
\end{minipage}

\end{figure*}

\section{The Common Pipeline Library}

The Common Pipeline Library (CPL) has been produced by ESO to aid in the 
development of VLT pipelines with standardised implementations and lower 
maintenance costs.
CPL is offered to instrument consortia and the user community to build 
instrument data reduction software. It is distributed as a development 
kit together with documentation and recipe front-ends. It is available on 
the web at 
\href{http://www.eso.org/cpl}{\texttt{www.eso.org/cpl}}.


A CPL based pipeline is structured as a set of recipe plug-ins which
are called by front-end applications (EsoRex or Gasgano). 
The recipes of the Data Reduction Software (DRS) are implemented 
as CPL plug-ins - units of code which are dynamically loaded into a host 
application at run time. The parent application is shielded from the recipe 
plug-in's implementation details by the CPL plug-in interface. This interface
consists of three functions, to initialize, execute and destroy the recipe
plug-in. These functions need to be provided by the recipe implementation.

The SINFONI pipeline is the CPL implementation of the Python based 
{\bf SPRED} package, the SPIFFI data reduction software developed by the 
Max-Planck-Institut  f\"ur extraterrestrische Physik  
(MPE, see also \cite{abuter06} and \cite{schreiber04}).

\section{Pipeline Front-End Applications}
Gasgano is a data management tool that simplifies the data organisation 
process. It offers automatic data classification and association
(Fig. \ref{fig:Gasgano}).
Gasgano classifies the raw and reduced data files based on their FITS 
keyword content by applying instrument specific rules.
It allows users to execute recipes directly from a GUI with flexible data 
reduction strategies. Gasgano is available from  the ESO web pages 
at \href{http://www.eso.org/gasgano}{\texttt{www.eso.org/gasgano}}.

EsoRex, the ESO Recipe executor, allows the execution of recipes from the 
command line. It does not offer the high level functionalities of Gasgano, 
but it can easily be embedded in scripts.

\section{Integral Field Spectroscopy}
\begin{figure*}[ht]
\centering
\includegraphics*[width=8truecm]{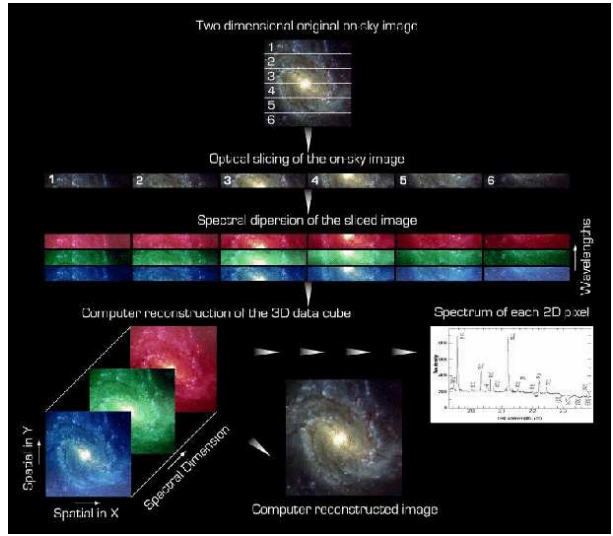}
\caption{This figure describes the Integral Field Spectroscopy concept: 
a two dimensional image on the sky is sliced, dispersed and imaged on the 
detector. The data reduction software then reconstructs a cube 
containing the full spatial and spectral information.\label{fig:IFS}}
\end{figure*}

For SINFONI, the instrument's two dimensional field of view (FOV) is sliced 
into a one dimensional pseudo long-slit, which is then dispersed by a 
spectrometer onto the two dimensional detector array (see Fig. \ref{fig:IFS} 
for a simplified overview of the principle). The recorded data contains
 the spatially resolved image slices (in one direction) and 
 the spectra (in the other).
The data reduction software has the task of removing the instrument signature,
in this case reconstructing the full spatial and spectral information into
a 3D data cube. Here, the FOV is mapped into the first and second dimensions, 
with the wavelength in the third dimension.   

\section{SINFONI}

\begin{figure}[ht]
\centering\begin{center}
\begin{tabular}{cc}
{\includegraphics*[width=5.8truecm]{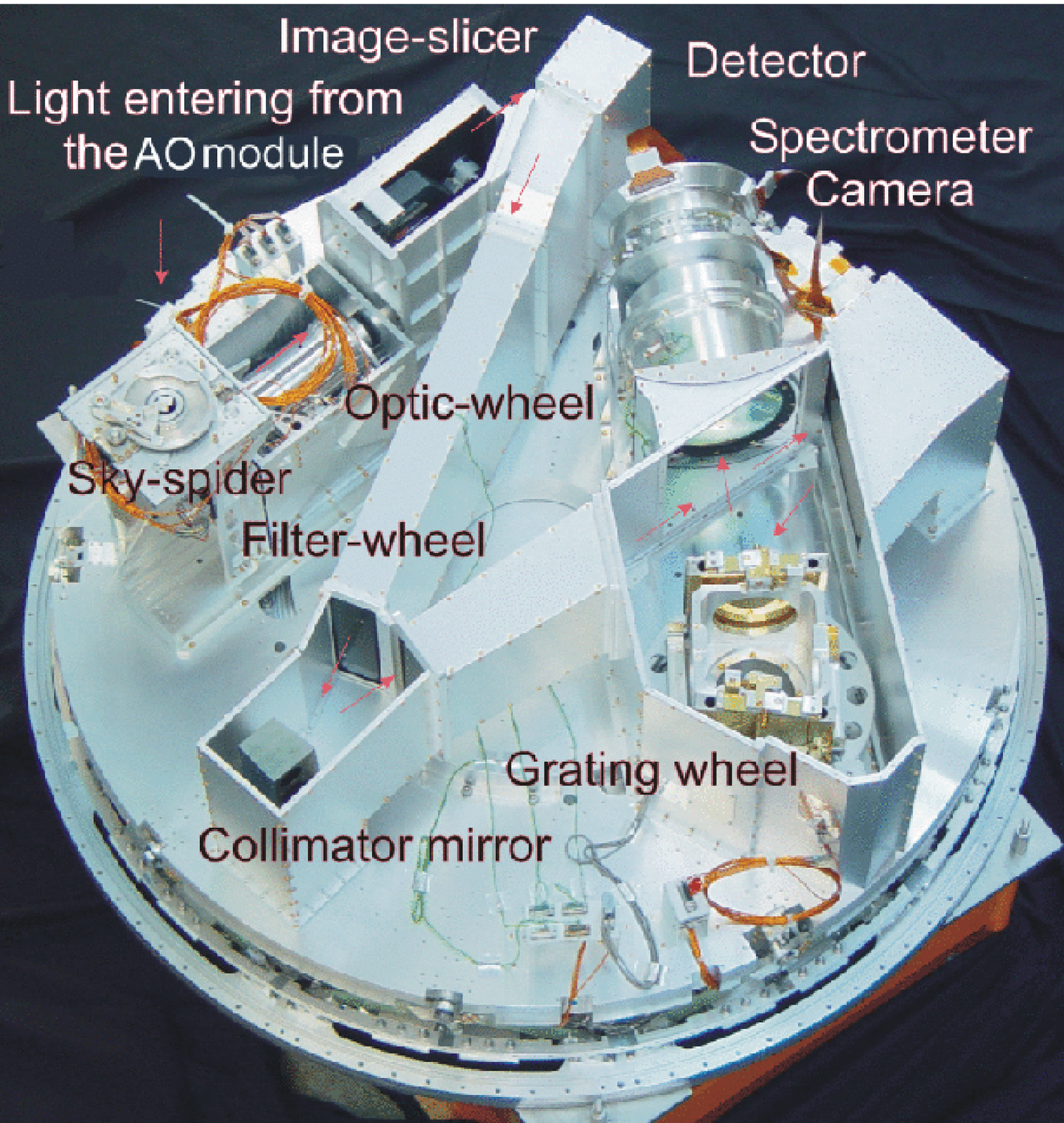}} &
{\includegraphics*[width=6.2truecm]{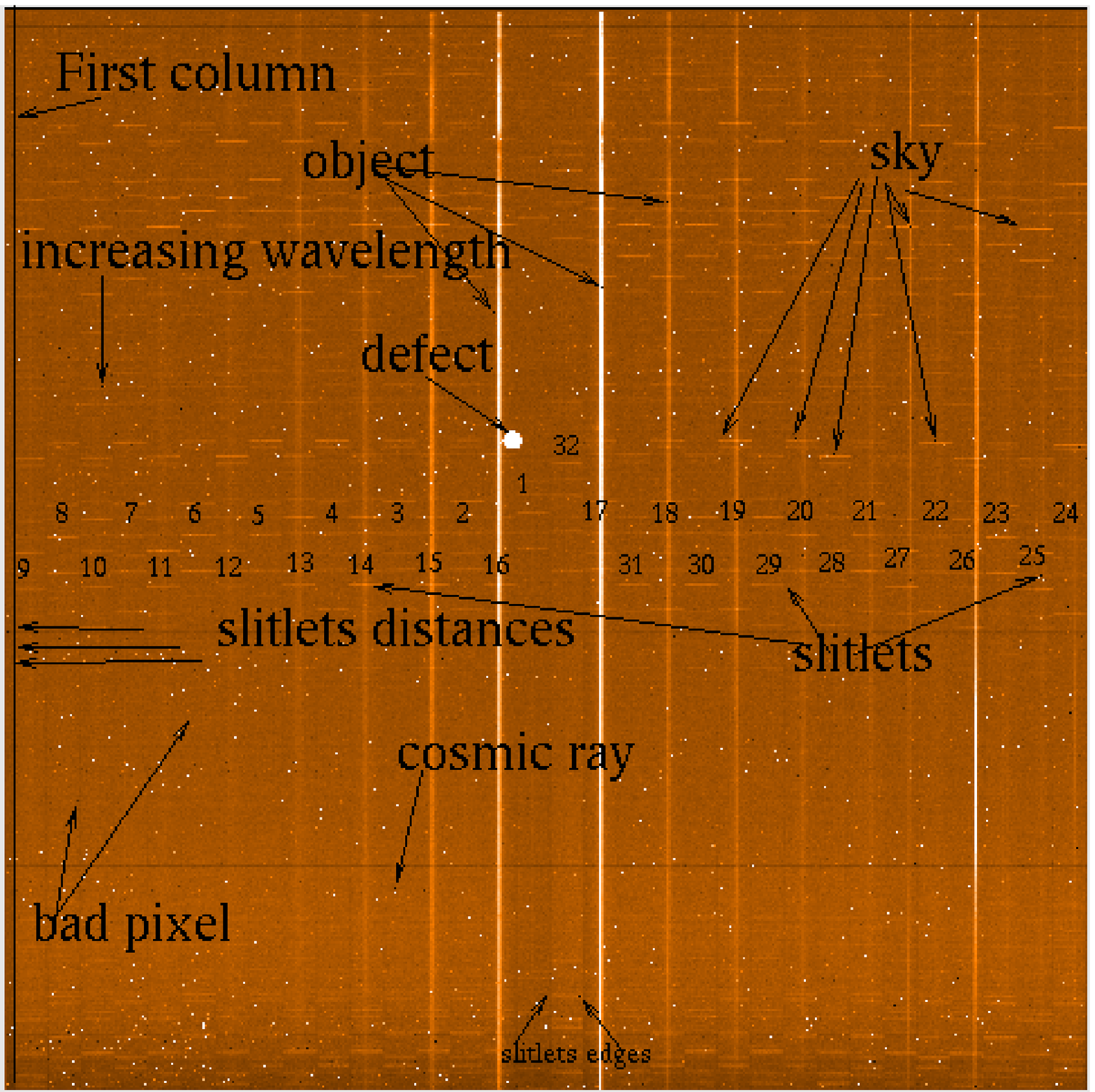}} \\
\end{tabular}
\end{center}
\begin{minipage}[t]{6.3truecm}
\caption{The SPIFFI spectrograph.\label{fig:SPIFFI}} 
\end{minipage}
\hfill
\begin{minipage}[t]{6.7truecm}
\caption{Typical data features.\label{fig:data_layout}} 
\end{minipage}
\end{figure}

SINFONI is composed of two subsystems: the Multi-Application Curvature 
Adaptive Optics (MACAO) unit, which allows diffraction and seeing limited 
observations, and SPIFFI, the SPectrometer for Infrared Faint Field Imaging, 
an integral field spectrograph which is shown in Fig. \ref{fig:SPIFFI}. 
It is described with more details in \cite{bonnet04} and \cite{eisenhauer03}.

Following the optical path, the light coming from the MACAO module is 
collimated by a doubled lens onto a cold stop to suppress the thermal 
background from the telescope secondary mirror.  Then the light path continues 
through a broadband filter (to suppress unwanted diffraction orders of the 
grating) and the optics wheel hosting the camera lens systems 
for the three different image scales (250, 100, 25 mas/pixel).
The images are focused on the small image slicer, a stack of 32 plane 
mirrors which slice the image into slitlets and separates the light from 
each slitlet into different directions. A second set of 32 mirrors, 
the big slicer, collects the light and forms a pseudo-slit which is then 
dispersed by the selected grating (J, H, K and HK bands) and finally imaged 
by the camera to the 2D NIR detector array.

The instrument FOV is rearranged on the detector into 32 vertical strips, 
(slices) each 64 pixels wide and 2048 pixels long, according to a given 
sequence in a brick-wall pattern. 
For SINFONI, the wavelength increases from the top to the bottom of the
detector array. 
Sky lines appear as small horizontal stripes, whilst cosmic rays and hot 
pixels manifest themselves as bright random spots (pixel aggregates). 
Dead pixels appear as black spots. Fig. \ref{fig:data_layout} displays many 
of these features.

\section{Data Reduction Cascade}

The SINFONI data reduction is described in more detail in the SINFONI Pipeline
User Manual, available on the web 
(\href{http://www.eso.org/pipelines}{\texttt{www.eso.org/pipelines}}).
An overview of the data reduction cascade is shown in Fig. \ref{fig:cascade}.
The following recipes are used in the data reduction process:
\begin{figure*}[ht]
\centering
\includegraphics*[width=8truecm]{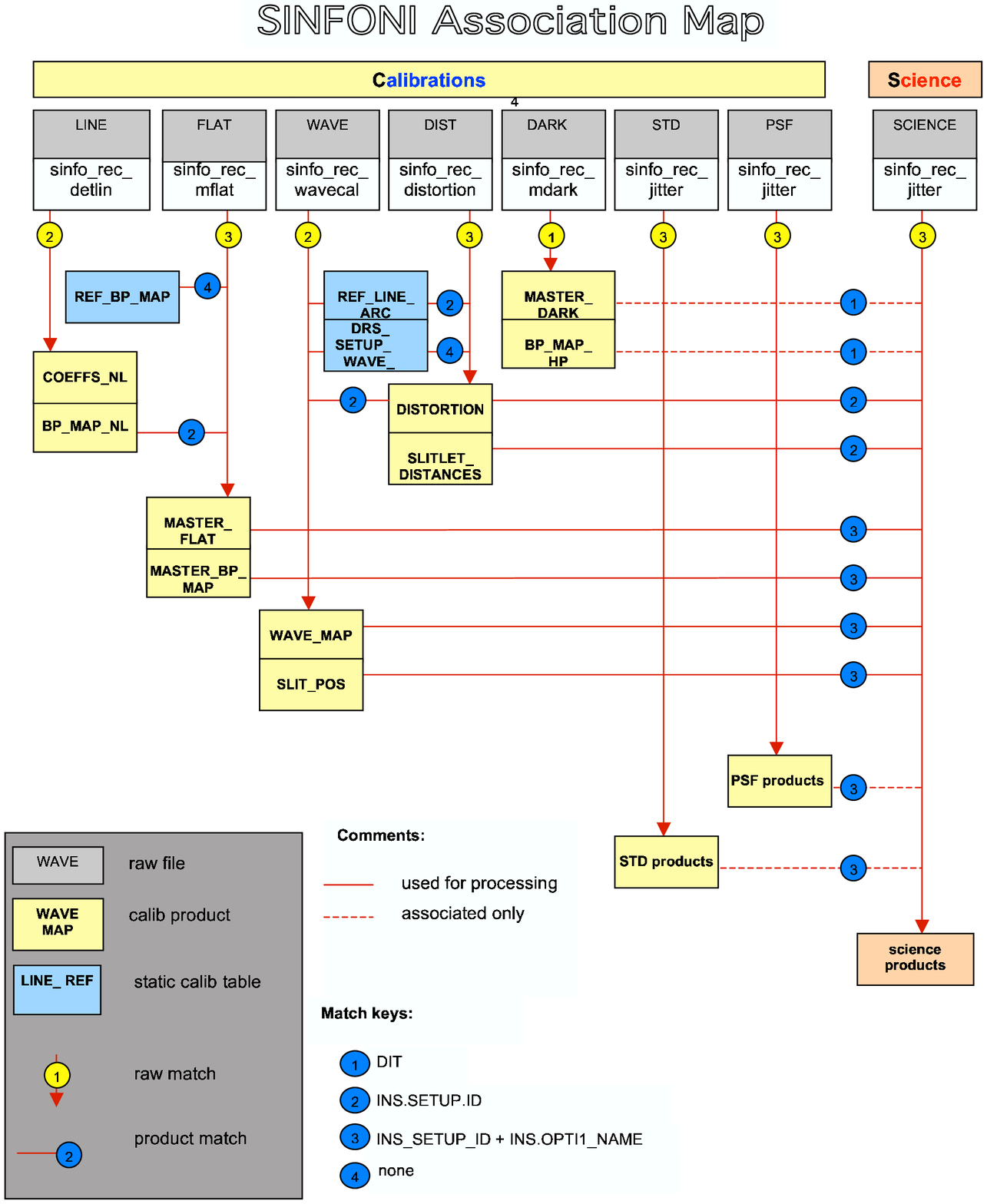}
\vspace{-2truecm}
\caption{Data reduction 
sequence with recipe and calibration product dependencies.\label{fig:cascade}}
\end{figure*}

\begin{itemize}
\item The linearity recipe derives a map of non-linear pixels from a 
  set of flat fields with increasing intensity.
\item The master dark recipe derives a master dark frame and a map of hot
  pixels from a set of dark frames.
\item The master flat recipe derives a master flat field frame and a map of 
  pixels whose intensity is above a given threshold from a set of flat field 
  frames. It combines this map with 
  the input non-linear pixel map and the reference bad pixel map to build 
  the master bad pixel map. 
\item The distortion recipe determines the optical 
  distortion coefficients for image reconstruction and the slitlets' 
  distance tables from a set of (typically 75) fibre flats. Fibre flats are 
  taken with the calibration fibre moved perpendicularly to the 
  image slices. Additional inputs are a pair of on/off arc lamp frames, 
  standard flats, a reference arc line list and a ``setup'' table.
\item The wavelength calibration recipe determines the wavelength map and 
  the slitlet position table from: a set of arc lamp frames, a master flat, 
  a master bad pixel map, a distortion table, a reference arc line table, 
  and a ``setup'' table. 

\item The jitter recipe reduces standard star calibrators and  
  science target observations at the bottom end of the data reduction 
  cascade. It requires as input a master bad pixel map, a master flat, 
  a master dark of the proper detector DIT, a distortion table, 
  a wavelength map, a slitlet distances table and a slitlet position table. 
  The recipe extracts the object spectrum (for sources whose Point Spread 
  Function is well approximated by a 2D Gaussian), 
  computes the instrument's Strehl ratio, determines instrumental throughput 
  and reconstructs a 3D cube. This cube is composed of a stack of planes,
  each of which contains the full spatial information of the instrument FOV.
  The wavelength information is mapped on the cube's third dimension. 
  If the recipe is provided with several object frames - for example, 
  a sequence of on/off-sky or jittered images - it will generate the sky 
  subtracted mosaic data cube from the raw data frames 
  (Fig. \ref{fig1:messineo}).

\end{itemize}

\section{Pipeline challenges}

The main tasks of the SINFONI pipeline are: 
the calculation of the optical distortion coefficients, 
the computation of a wavelength calibration solution,
the reconstruction of the full spatial and spectral information on a 3D data
cube, and the removal of the emission from the telescope, the instrument 
and the sky (dominant in the NIR).
Additionally, the pipeline should generate the information needed  
to monitor the behavior of the instrument as atmospheric conditions change. 
These QC1 parameters are used by DFO to monitor the quality of master 
calibrations, and science products; they also give an indication of the 
instrument's health (see \cite{hummel06}
for a more comprehensive overview). 

\subsection{Distortion Computation:}
The distortion function, which characterises the curvature of the spectral
traces of the 32 slices, is calculated from a series of continuous light 
fibre spectra taken with the fibre moved perpendicularly to the slices.
For every position of the fibre a raw FITS data file is recorded. 
With an initial data reduction step the single fibre spectra are co-added to
an all-fibers synthetic frame and the offset of the first slitlet is
calculated as a reference position. The distortion is computed in three
further steps:
\begin{itemize}
\item Detecting the fibres and tracing the curved fibre spectra.
\item Constructing two numerical grids on the distorted and the undistorted 
  fibre positions.
\item Solving a 2D polynomial fit to transform positions from raw to 
  undistorted coordinates.
\end{itemize}
Finally, the recipe determines the position of the edges of the slitlets 
by fitting the edges of the brightest arc line of each slitlet 
with a linear step function or a Boltzmann function.

\begin{figure}[ht]
\centering
\begin{center}
\begin{tabular}{cc}
{\hspace{0.5truecm}\includegraphics*[width=5truecm]{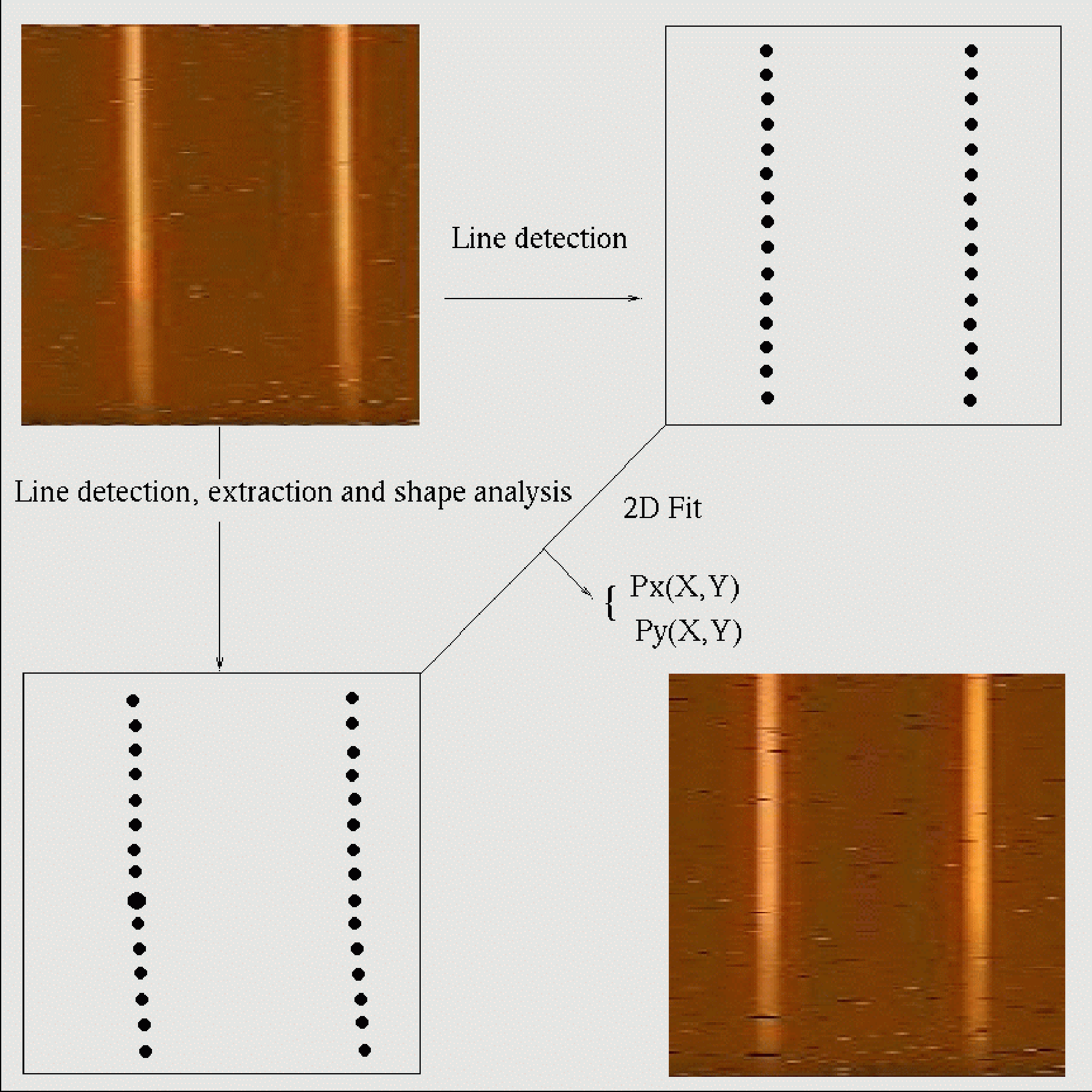}\hspace{1.0truecm}} &
{\hspace{1.0truecm}\includegraphics*[width=5truecm]{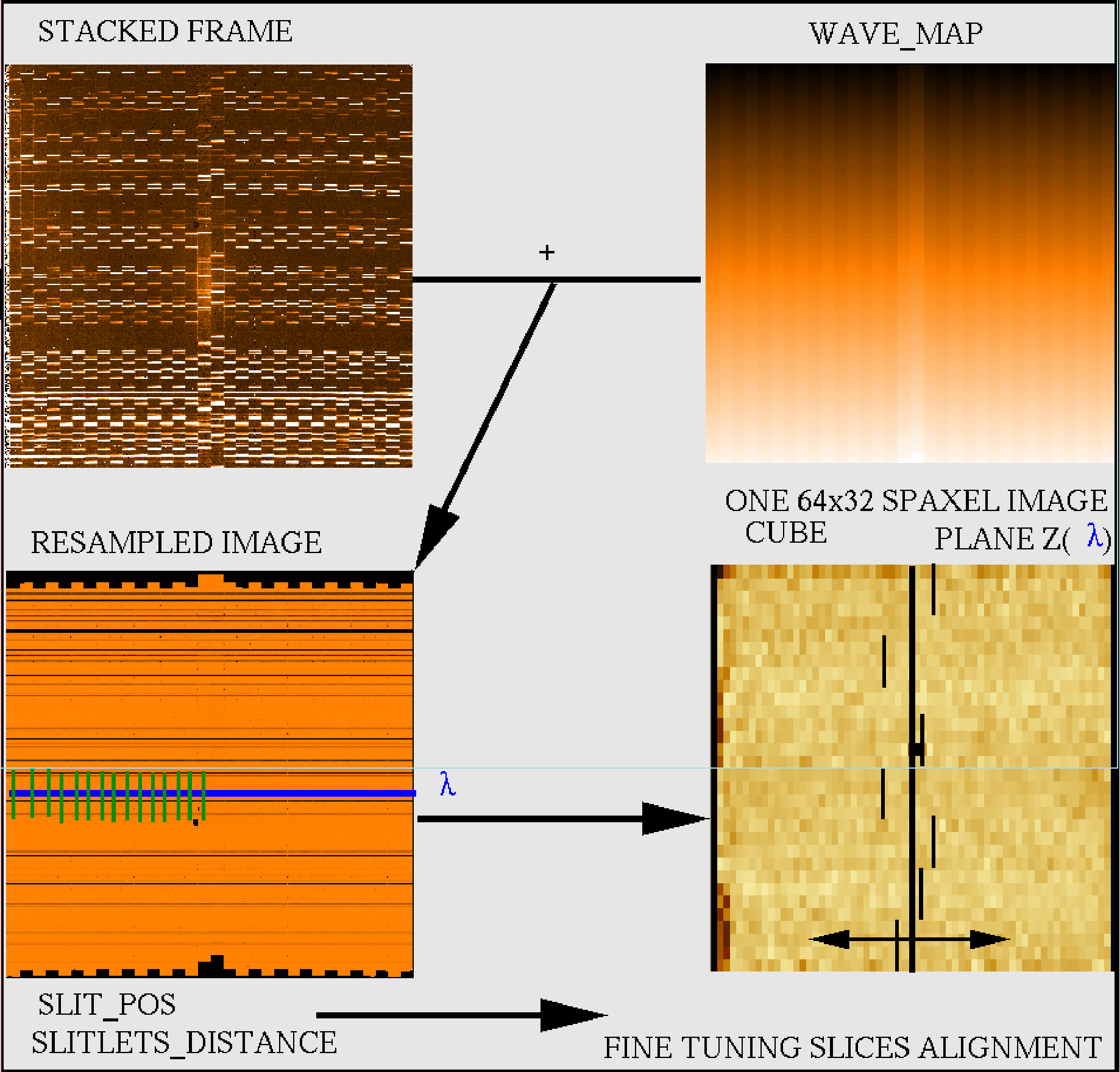}\hspace{0.5truecm}} \\
\end{tabular}
\end{center}
\begin{minipage}[t]{6.5truecm}
\caption{This image describes the concept of the distortion computation
algorithm. The spectra of the fibres are traced, two grids on the 
distorted and undistorted space are built, then a 2D polynomial transformation 
is performed. This figure displays only two of the 32 slitlets.\label{fig:DISTORTION}}
\end{minipage}
\hfill
\begin{minipage}[t]{6.5truecm}
\caption{Cube reconstruction: raw data are resampled using a wavelength map 
to remove the brick-wall pattern. The slitlets are then stacked into a cube 
taking slitlet distances and edge positions into account. 
Each plane of the cube is a monochromatic image of the instrument FOV.
\label{fig:3DCUBE}}
\end{minipage}
\end{figure}

\subsection{Wavelength Calibration:}
The wavelength calibration is based on a set of on/off arc lamp exposures.
Initially, the arc lamp frames are stacked and the off-frames are subtracted 
from the on-frames to remove the thermal background. 
The resulting frame is flat fielded, the
distortion correction is applied, and static bad pixels are corrected. 

Using a polynomial model, guess values for the central wavelength 
$\lambda_{0}$ and the first and second order dispersion coefficients, 
it is possible to build a synthetic frame from the reference arc line list 
by associating at each catalog entry a delta-like function of proper wavelength
and intensity. Then this spectrum is convolved 
with a Gaussian having a FWHM appropriate to the instrumental resolution. 
The synthetic frame thus has a well known wavelength and intensity.
Cross correlation of the synthetic frame with the observed spectrum provides
an approximate wavelength and line intensity for each raw (and emission line).

Doing a non-linear least squares fit with a Gaussian to the synthetic frame,
one can associate to each line an accurate intensity, position, FWHM and 
background. This permits the construction of line index triplets 
$(x,y,\lambda)$.
After removal of false identifications, dispersion coefficients are derived 
by a polynomial fit along each detector's column.
Finally the smoothed wavelength dispersion coefficients are obtained with a 
low order polynomial interpolation of each dispersion coefficient in the 
detector's spatial direction.

\subsection{3D Data Cube Reconstruction:}
\begin{figure*}[th]
\centering
\begin{center}
\begin{tabular}{cc}
{\includegraphics*[width=5.5truecm]{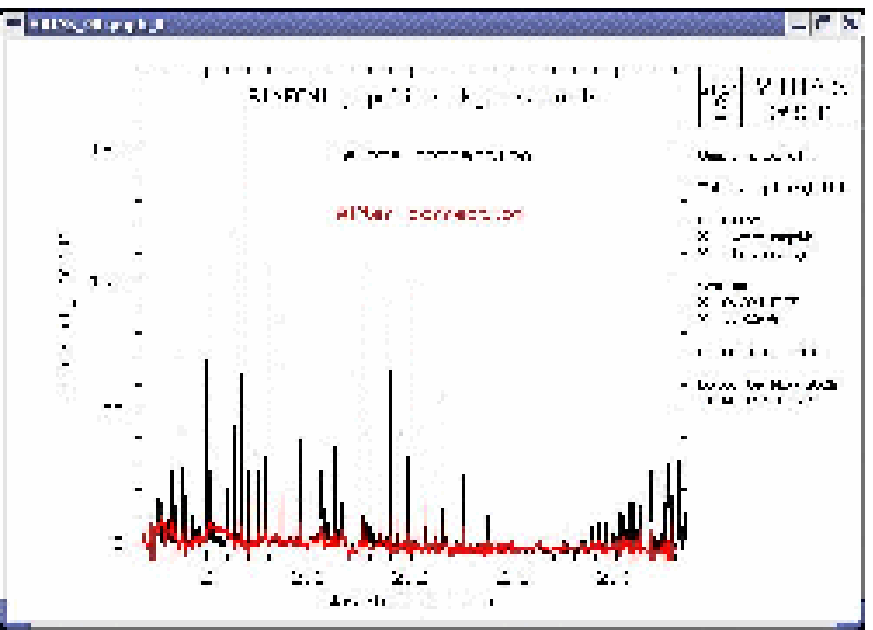}} &
{\hspace{0.5cm}\includegraphics*[width=7.0truecm]{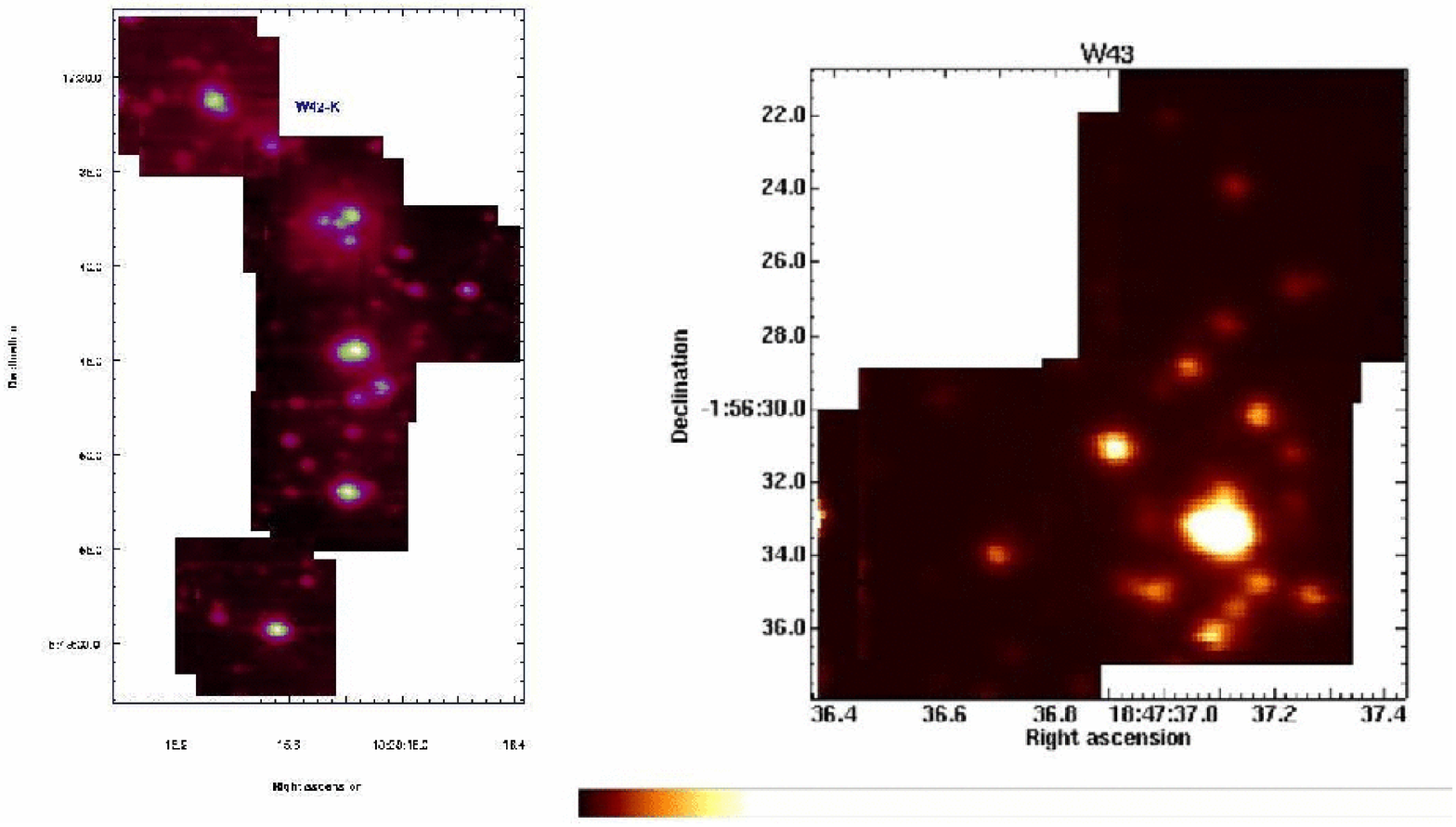}} \\
\end{tabular}
\end{center}
\begin{minipage}[t]{5.5truecm}
\caption{Comparison between un-corrected (black) and corrected (red)
sky subtracted object spectra.\label{fig:sky-problem}}
\end{minipage}
\hfill
\begin{minipage}[t]{7.5truecm}
\caption{Observations of stars embedded in the giant HII regions W42 (left) 
and W43 (right).
SINFONI allows to resolve the cluster and extract the spectra of many stars.
\label{fig1:messineo}}
\end{minipage}

\end{figure*}

From the input set of frames  object-sky pairs are defined by associating to
each object frame the sky frame that is closest in time. 
For each object-sky pair a 3D data cube is constructed as follows.
If a master dark is provided as input of the jitter recipe, 
this is subtracted from the object and the sky frames, 
then the corrected sky frame is subtracted from the corrected 
object frame, and the resulting frame is flat fielded and distortion 
corrected. 
The resulting image is re-sampled into the same wavelength steps, a process 
which aligns the different slitlets in wavelength. A 3D data cube is 
subsequently constructed using the information read from the slitlet 
position table and the slitlet edge distances table. As not all of the 
slitlets are exactly 64 pixels long and 64 pixels distant from the adjacent 
one, it is necessary to refine the slitlets' position alignment on the stacked 
cube. 
The jitter recipe then constructs a mosaic of the series of input
data cubes (each aligned in the co-added space) by reading the 
cumulative offset information from the FITS headers of the raw input frames
(Fig. \ref{fig1:messineo}).
Finally, the recipe extracts the spectrum along the Z-axis.






\subsection{Sky subtraction}


In its most recent release the pipeline corrects for improper sky subtraction;
in older releases this caused residuals in on-off sky-subtracted frames. 
The sky emission is removed from the object observation by 
subtracting the sky frame from the object frame. This assumes that the two 
frames have a stable spectral format.
However, it has been found that in on-off, object-sky sequences the 
instrument setting can occasionally have instabilities up to 
a significant fraction of a pixel. The residual sky features
(resembling a P-Cygni profile) can be significantly reduced by using 
an improved data reduction procedure developed in collaboration with 
Ric Davies from MPE (Fig. \ref{fig:sky-problem} and \cite{davies07}).



\section{Science Results}

SINFONI started science operations on April 1, 2005, and has since then
obtained science data for over 1000 hours of science integration time - 
not including acquisition overheads, calibrations or standard star 
observations.
The scientific observations discussed below were taken within the science 
verification runs before the start of operations and reduced with the SINFONI 
pipeline. The results are given as examples for the
performance of the instrument and the pipeline data reduction.

\cite{messineo06} have observed stars embedded in the stellar clusters 
of the giant HII regions W42 and W43 (Fig. \ref{fig1:messineo}).
In W42 they have resolved the cluster and extracted the spectra of 7 stars 
with high S/N of $>$30 in H and K bands. This has doubled the number of 
spectral detections with respect to what was available in the literature. 
In W43 the cluster centre has been resolved, and they have detected the 
photospheric lines of several of the 13 brightest stars (9 stars in K and 4 
in H band).

\begin{figure}[th]
\centering
\begin{center}
\begin{tabular}{cccc}
{\includegraphics*[width=5.5truecm]{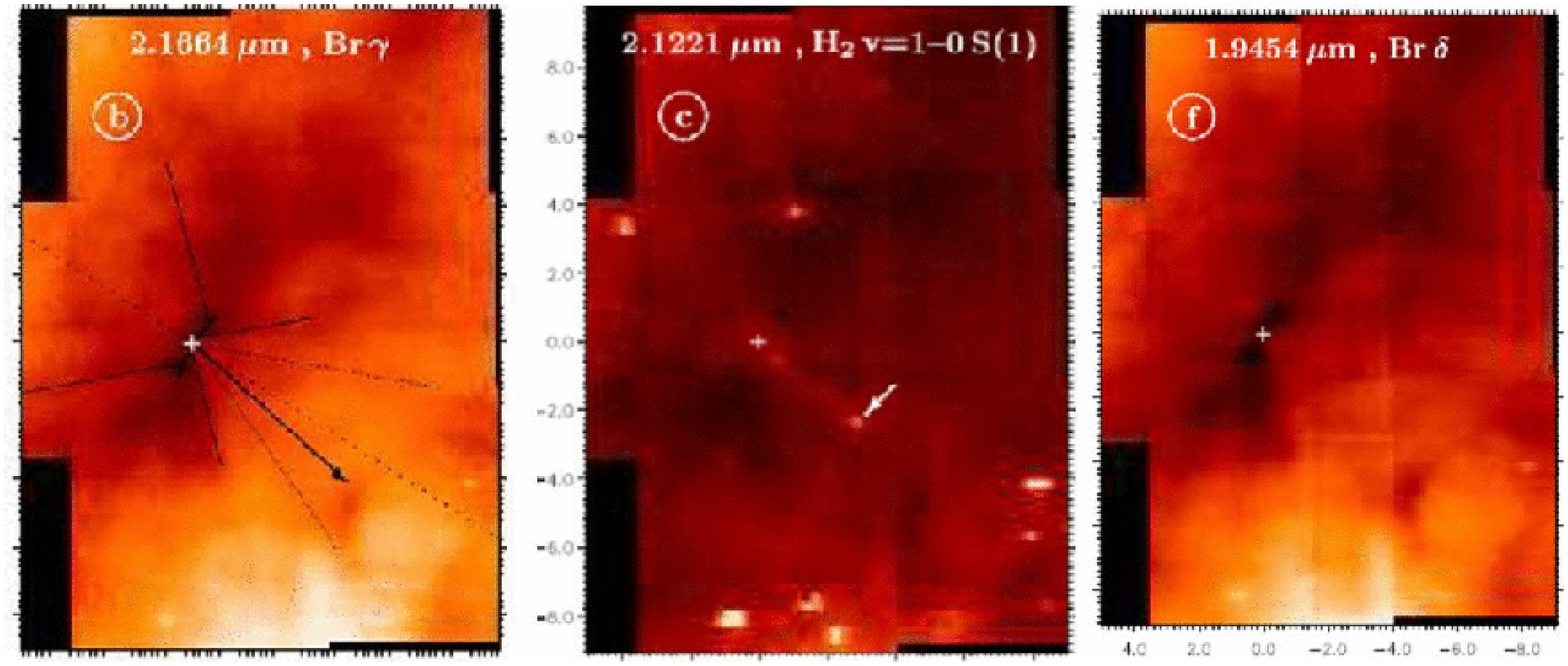}} &
{\includegraphics*[width=7.5truecm]{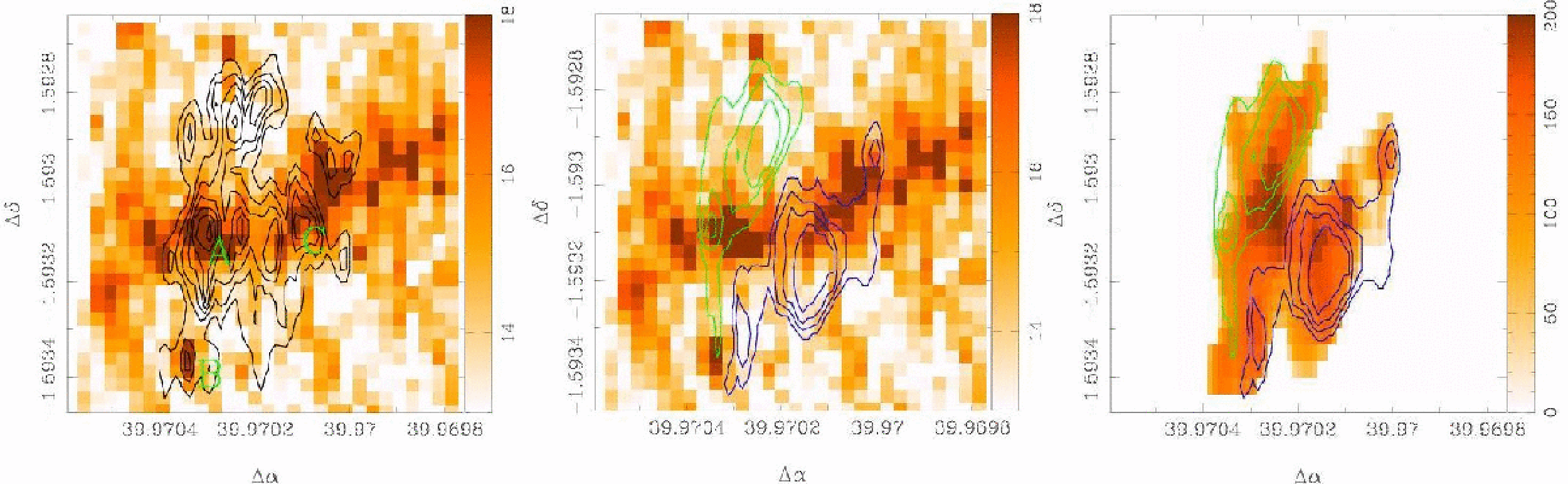}} &
\end{tabular}
\end{center}
\begin{minipage}[t]{5.5truecm}
\caption{The discovery of a collimated H$_{2}$ jet in M17 (center) is an 
indication of ongoing accretion processes. On the left side is indicated a 
possible geometry of the jet. On the right it is shown the Br$_{\delta}$ map.
\label{fig:nurnberger}}
\end{minipage}
\hfill
\begin{minipage}[t]{7.5truecm}
\caption{Star formation in a high-red shift galaxy (A370-A5). 
Observations suggest that it is not a rotating disk, and that it is 
probably characterised by a bipolar outflow.
\label{fig:lemoine}}
\end{minipage}
\end{figure}

\cite{nuernberger07} were studying the formation of massive proto-stars 
through disk accretion of gas in M17. From the SINFONI H$_{2}$ maps they
report the discovery of an H$_{2}$ jet which apparently arises from 
the suspected proto-stellar source(s) located at the very center of the disk. 
They can infer the diameter and sub-structure of the innermost part of the 
flared disk from the Br$_{\gamma}$, Br$_{\delta}$ and HeI maps.
As the ejection of material through a jet and/or the outflow is always 
linked to accretion of gas and dust onto the circumstellar disk or onto the 
central proto-stellar source(s), the presence of a collimated H$_{2}$ jet 
provides direct and unquestionable evidence for ongoing accretion processes 
in the case of this disk.

\cite{lemoine07} (see also \cite{gillessen06})
used SINFONI to probe the star formation of the high-red shift galaxy 
A370-A5. 
The clumpy structure evidenced by the H$_{\alpha}$ contour plots are an 
indication of a stellar component and show two extended regions without 
continuum counterpart (Fig. \ref{fig:lemoine}, left).
H$_{\alpha}$ velocity maps indicate the presence of a velocity 
gradient perpendicular to the stellar emission (Fig. \ref{fig:lemoine}, 
center). 
The H$_{\alpha}$ velocity dispersion map shows no indication of the central 
peak which would characterise a rotating disk (Fig. \ref{fig:lemoine}, right). 
From these observations Lemoine-Busserolle et al. conclude that A370-A5 is
not a rotating disk, and it is probably characterised by a bipolar outflow.


\section{Summary}

The SINFONI pipeline is an important part of the ESO-VLT Data Flow System
to support SINFONI operations, monitor the instrument's health, generate 
master 
calibration and scientific products, and assess their quality.
Applying the recently improved sky subtraction, the master dark, master flat
and optical distortion corrections, and the wavelength calibration, a good
quality 3D data cube is constructed, which contains the full spatial and 
spectral information of the SINFONI FOV for further science analysis.


\vspace{-1truecm}
\section*{\bf{Acknowledgments}} \smallskip

We would like to thank the SINFONI Instrument Operation Team for the 
feedback provided during commissioning and early phases of 
operations, which helped to improve the quality of the data reduction.
We kindly acknowledge the provision and discussion of reduced 
science data by the science verification team members: Maria Messineo, 
Marie Lemoine-Busserolle and Dieter N\"urnberger.
We appreciate the support and suggestions provided by the ESO CPL and 
pipeline team to improve the SINFONI pipeline.


\vspace{-1truecm}
\begin{thebibliography}{8}
\bibitem[Abuter et~al.(2006)]{abuter06}
Abuter R., Schreiber J., Eisenhauer F., Ott T., Horrobin M., Gillesen S., {\em SINFONI data reduction software}, 2006, New Astronomy Reviews, {\bf 50}, Issue 4-5, 398, 2006.

\bibitem[Schreiber et~al.(2004)]{schreiber04}
Schreiber J., Thatte N., Eisenhauer F., Tecza M., Abuter R. \& Horrobin M.,
 {\em Data Reduction Software for the VLT Integral Field Spectrometer SPIFFI}
 2004 in ASP Conf. Ser., Vol. 314, 380 - 383, Astronomical Data Analysis Software and Systems XIII,
 eds. Francois Ochsenbein, Mark G. Allen and Daniel Egret.


\bibitem[Ballester et~al.(2006)]{ballester06}
Ballester P. et~al., {\em Data reduction pipelines for the Very Large Telescope}, 2006, Proceedings of the SPIE, 6270, 2006.

\bibitem[Bonnet et~al.(2004)]{bonnet04}
Bonnet H. et~al., {\em First Light of SINFONI at the VLT}, 2004, 
The Messenger, {\bf 117}, 17.

\bibitem[Davies R., (2007)]{davies07}
Davies R., {\em A method to remove residual OH emission from near infrared spectra}, 2007, Monthly Notices Reviews Astronomical Society, accepted.

\bibitem[Eisenhauer et~al. (2003)]{eisenhauer03}
Eisenhauer F., et~al. {\em SINFONI - Integral field spectroscopy at 50 milli-arcsecond resolution with the ESO VLT}, 2003, SPIE, {\bf 4841}, 1548.

\bibitem[Gillessen et~al. (2006)]{gillessen06}
Gillessen S., et~al., 
{\em First Science with SINFONI}, 2006, The Messenger, {\bf 120}, 26.

\bibitem[Hummel et~al.(2006)]{hummel06}
Hummel W., Modigliani A., Dumas C., Amico P., Szeifert T., Neeser M., {\em Quality Control of the 3D spectrograph SINFONI}, 2006, These Proceedings.

\bibitem[Lemoine-Busserolle et~al.(2007)]{lemoine07}
Lemoine-Busserolle M., Sanchez S.F., Kissler-Patig M., Pell\'o R., Kneib J.-P., Bunker A., Contini T., 2007, A\&A, submitted. 

\bibitem[Messineo et~al. (2006)]{messineo06}
Messineo M., Petr-Gotzens M., Menten K.M., Sculler F. \& Habing H.J. {\em Galactic structure and stellar clusters in obscured \hbox{\sc Hii} regions}, 2006, Proceedings of the Galactic Center Workshop ``From the Center of the Milky Way to Nearby Low-Luminosity Galactic Nuclei''. Eds Reiner Schoedel.

\bibitem[N\"urnberger et~al.(2007)]{nuernberger07}
N\"urnberger D.E.A., Chini R., Eisenhauer F., Kissler-Patig M., Modigliani A., Siebenmorgen R., Sterzik M.F., Szeifert T., {\em Formation of a massive protostar through disk accretion of gas}, 2007, A\&A, submitted. 

\bibitem[Silva D. \& Peron M. (2004)]{silva05}
Silva D. \& Peron M., {\em VLT Science Products Produced by Pipelines},
2004, The Messenger {\bf 118}, 2.
\end{thebibliography}


\clearpage
\end{document}